\documentclass[12pt,a4paper]{article}
\usepackage{amssymb}
\usepackage[dvips]{graphicx}
\usepackage{bm}
\usepackage{amsmath}
\begin{document}

\title{ Exact results in explicit three-loop calculations using higher derivatives for ${\cal N}=1$ SQCD}

\author{A.L. Kataev$^{a,b}$, A.E. Kazantsev$^{c,*}$, K.V. Stepanyantz$^c$\\
{\small{\em $^a$Institute for Nuclear Research of the Russian Academy of Sciences,}}\\
{\small{\em 117312, Moscow, Russia}}\\
{\small{\em $^b$Moscow Institute of Physics and Technology,}} 
{\small{\em 141700 Dolgoprudnyi, Russia}}\\
{\small{\em $^c$Moscow State University}}, {\small{\em  Physical Faculty, Department  of Theoretical Physics}}\\
{\small{\em 119991, Moscow, Russia}}\\
{\small{\em $^*$e-mail: kazancev@physics.msu.ru}}}

\date{}

\maketitle
\vspace{-2em}
\begin{abstract}
We calculate the three-loop Adler $D$-function of ${\cal N}=1$ SQCD regularized by higher covariant derivatives and find the subtraction scheme in which the exact NSVZ-like relation for this function proposed in Refs. \cite{Shifman:2014cya,Shifman:2015doa} is valid.
\end{abstract}

\vspace*{-11.5cm}

\begin{flushright}
INR-TH-2017-029
\end{flushright}

\vspace*{10.0cm}

\section{Introduction}

Recently it was proposed \cite{Shifman:2014cya,Shifman:2015doa} that the Adler $D$-function of ${\cal N}=1$ SQCD is related to the anomalous dimension of the chiral matter superfields by the equation
\begin{equation}\label{ShifmanFormula}
D(\alpha_{s0})=\frac{3}{2}N\sum\limits_{\alpha=1}^{N_{f}} q_{\alpha}^2\Big(1-\gamma(\alpha_{s0})\Big),
\end{equation}
\noindent
where $N$ is the number of colors,  $N_{f}$ is the number of quark flavors, and $q_\alpha$ are quark electric charges. This relation is very similar to the NSVZ exact $\beta$-function \cite{Novikov:1983uc}. Eq. (\ref{ShifmanFormula}) was proved for the $D$-function and anomalous dimension defined in terms of the bare coupling constant, which are independent of the subtraction scheme. However, for extracting physical consequences it is necessary to use the language of finite, renormalized couplings, which implies a choice of a particular subtraction scheme. However, the $D$-function and the anomalous dimension are scheme-dependent from the three- and two-loop order respectively, and Eq. \eqref{ShifmanFormula} is not in general satisfied. Thus, it is necessary to find the subtraction scheme in which Eq. \eqref{ShifmanFormula} is valid for the renormalization group functions defined in terms of the renormalized charges. We argue that such a scheme exists and is fixed by the condition that the renormalization constants should be equal to unity at some fixed value $x_0$ of $\ln(\Lambda/\mu)$ in the case of using the higher covariant derivative regularization. (Here $\Lambda$ is the cut-off parameter and $\mu$ is the renormalization scale.) We will see that for $x_0=0$ this prescription is equivalent to subtracting exclusively powers of $\ln(\Lambda/\mu)$. This is the reason why we call it $\mbox{HD}+\mbox{MSL}$, where HD is the abbreviation of Higher Derivatives, and MSL stands for Minimal Subtraction of Logarithms.

Until now the $D$-function of ${\cal N}=1$ SQCD has been known to the order  $\alpha_s$  \cite{Kataev:1983at,Altarelli:1983pr}. Here, using the higher covariant derivative regularization, we calculate it to the order $\alpha_s^2$, where it becomes scheme-dependent. First, we do this in terms of the bare coupling and verify Eq. \eqref{ShifmanFormula} to the three-loop order. Then, we reformulate the results in terms of the renormalized coupling constant and check the $\mbox{HD}+\mbox{MSL}$ prescription for the NSVZ-like scheme in this approximation.

\section{The Adler D-function to three-loop order}

We consider the ${\cal N}=1$ supersymmetric Yang--Mills theory with $N_{f}$ pairs of chiral superfields $\phi_{\alpha}$ and $\widetilde{\phi}_{\alpha}$ in a representation $R$ and its complex conjugate of a simple group $G$ (with the bare coupling $g_0$) and with the charges $q_\alpha$ and $-q_\alpha$ with respect to $U(1)$ (with the bare coupling $e_0$). We insert into the action powers of higher covariant derivatives and, to cancel one-loop divergences, we introduce a set of three commuting chiral superfields in the adjoint representation of $G$, with the masses proportional to $\Lambda$ with the coefficient $a_{\varphi}$, and also  a set of anticommuting chiral superfields $\Phi_{\alpha}$ and $\widetilde{\Phi}_{\alpha}$, $\alpha=1,...,N_{f}$, in the same representations of the group $G$ and with the same $U(1)$ charges as $\phi_{\alpha}$ and $\tilde{\phi}_{\alpha}$ respectively, with the mass proportional to $\Lambda$ with the coefficient $a$.

The renormalized coupling constants $\alpha_s=g^2/4\pi$ and $\alpha=e^2/4\pi$ and the renormalization costants for the chiral matter superfields $Z_i{}^j$, such that
$\phi_{\alpha i}=\sqrt{Z}_i{}^j(\phi_R)_{\alpha j}$ and $\widetilde{\phi}_\alpha^i=\sqrt{Z}_{j}{}^i(\widetilde{\phi}_R)_{\alpha}^j$, are defined in the standard manner. We can also define the renormalization constants for the couplings: $Z_{\alpha_s}=\alpha_s/\alpha_{s0}$ and $Z_{\alpha}=\alpha/\alpha_0$. The one-loop relation between the bare and renormalized strong coupling has the form
\begin{equation}\label{One-LoopRunning}
\frac{1}{\alpha_{s0}}-\frac{1}{\alpha_{s}}=\frac{1}{2\pi}\Big[3C_{2}\Big(\ln\frac{\Lambda}{\mu}+b_{11}\Big) - 2 N_{f} T(R) \Big(\ln\frac{\Lambda}{\mu}+b_{12}\Big)\Big] + O(\alpha_{s}),
\end{equation}
\noindent
where $b_{11}$ and $b_{12}$ are finite constants that depend on a choice of a subtraction scheme.

Let us write the definition of the $D$-function and the anomalous dimension in terms of the bare coupling constant
\begin{equation}\label{RGFunctionsBare}
D(\alpha_{s0})=-\left.\frac{3\pi}{2}\frac{d}{d\ln\Lambda }\alpha_{0}^{-1}\left(\alpha,\alpha_{s},\Lambda/\mu\right)\right|_{\alpha,\alpha_{s}=\mbox{\scriptsize const}};\qquad\gamma(\alpha_{s0})_i{}^j=-\frac{d(\ln Z)_i{}^j}{d\ln\Lambda}\biggr|_{\alpha_s=\mbox{\scriptsize const}},
\end{equation}
\noindent
and in terms of the renormalized coupling constant
\begin{equation}\label{RGFunctionsRenorm}
\widetilde{D}(\alpha_{s})=-\left.\frac{3\pi}{2}\frac{d}{d\ln\mu }\alpha^{-1}\left(\alpha_0,\alpha_{s0},\Lambda/\mu\right)\right|_{\alpha_0,\alpha_{s0}=\mbox{\scriptsize const}};\qquad \widetilde{\gamma}(\alpha_{s})_i{}^j=\frac{d(\ln Z)_i{}^j}{d\ln\mu}\biggr|_{\alpha_{s0}=\mbox{\scriptsize const}}.
\end{equation}

The calculations show that in the considered order of perturbation theory the $D$-function defined by Eq. \eqref{RGFunctionsBare} is given by an integral of a (double) total derivative with respect to the momentum of the matter loop. This can be calculated by picking up contributions of poles of the massless propagators. For the higher derivative regulator $R(x)=1+x^n$ the results for the $D(\alpha_{s0})$ in the three-loop approximation and $\gamma_i{}^j(\alpha_{s0})$ in the two-loop approximation have the form  \cite{Kataev:2017qvk}
\begin{eqnarray}\label{ThreeLoopDBare}
&& D(\alpha_{s0}) = \frac{3}{2} \sum\limits_{\alpha=1}^{N_{f}}q_{\alpha}^2 \Big[ \mbox{dim}(R) + \frac{\alpha_{s0}}{\pi} \mbox{tr}\,C(R) + \frac{3\alpha_{s0}^2}{2\pi^2}\, C_{2}\, \mbox{tr}\,C(R) \Big(\ln a_\varphi + 1\Big) \quad\nonumber\\
&& - \frac{\alpha_{s0}^2}{\pi^2}\, N_{f}T(R)\, \mbox{tr}\, C(R) \Big(\ln a  +1\Big) - \frac{\alpha_{s0}^2}{2\pi^2}\, \mbox{tr}\left( C(R)^2\right) \Big] + O(\alpha_{s0}^3)
\end{eqnarray}
\noindent
and
\begin{eqnarray}\label{TwoLoopGammaBare}
&& \gamma(\alpha_{s0})_i{}^j = - \frac{\alpha_{s0}}{\pi} C(R)_i{}^j - \frac{3\alpha_{s0}^2}{2\pi^2} C_2 C(R)_i{}^j  \Big(\ln a_\varphi + 1\Big) + \frac{\alpha_{s0}^2}{\pi^2} N_{f} T(R) C(R)_i{}^j \Big(\ln a + 1\Big)\nonumber\\
&& + \frac{\alpha_{s0}^2}{2\pi^2} \left(C(R)^2\right)_i{}^j  + O(\alpha_{s0}^3),
\end{eqnarray}
\noindent
where $\mbox{tr}(T^{A}T^{B})=T(R)\delta^{AB}$, $C(R)_i{}^j=(T^{A}T^{A})_i{}^j$, $C_{2}\delta^{CD}=f^{ABC}f^{ABD}$, and $\mbox{dim}(R)$ is the dimension of the representation $R$. It is easy to see that these functions satisfy the NSVZ-like relation
\begin{equation}\label{ThreeLoopRelation}
D(\alpha_{s0})=\frac{3}{2}\sum\limits_{\alpha=1}^{N_{f}} q_{\alpha}^2\Big(\mbox{dim}(R) - \mbox{tr}\, \gamma(\alpha_{s0})\Big) + O(\alpha_{s0}^3),
\end{equation}
\noindent
which gives \eqref{ShifmanFormula} when the gauge group is $SU(N)$ and the matter superfields belong to the fundamental (and antifundamental) representation.

\section{The NSVZ-like subtraction scheme}

Similarly to \cite{Kataev:2013eta},  we define the subtraction scheme in which we expect the relation \eqref{ShifmanFormula} to be valid for the RG functions defined in terms of the renormalized coupling constant by a set of boundary conditions imposed on the renormalization constants at some fixed value $x_0$ of $\ln(\Lambda/\mu)$,
\begin{equation}\label{Condition}
Z(\alpha_{s},x_{0})_i{}^j=\delta_i{}^j;\qquad Z_{\alpha}(\alpha,\alpha_{s},x_{0})=1;\qquad Z_{\alpha_{s}}(\alpha_{s},x_{0})=1.
\end{equation}
Moreover, in this scheme $\widetilde{D}$ and $\widetilde{\gamma}_i{}^j$ have the same expansion in $\alpha_{s}$ as respectively $D$ and $\gamma_i{}^j$ have in $\alpha_{s0}$ and, therefore, satisfy the equation similar to Eq. \eqref{ThreeLoopRelation} but with the argument $\alpha_s$ (see \cite{Kataev:2017qvk} for all details).

Integrating Eqs. \eqref{ThreeLoopDBare} and \eqref{TwoLoopGammaBare} with respect to $\ln\Lambda$ with the fixed renormalized couplings, we can recover the expressions for $\alpha_0^{-1}-\alpha^{-1}$ and $(\ln Z)_i{}^j$ respectively. The result contains some constants of integration, which are interpreted as finite constants fixing a subtraction scheme. Then, differentiating the obtained expressions with respect to $\ln\mu$ according to Eq. \eqref{RGFunctionsRenorm} we find the $D$-function defined in terms of the renormalized coupling,
\begin{eqnarray}\label{ThreeLoopDRenormalized}
&&\widetilde{D}(\alpha_{s}) =\frac{3}{2}\sum\limits_{\alpha=1}^{N_{f}}q_{\alpha}^2\biggl(\mbox{dim}(R) + \frac{\alpha_{s}}{\pi}\, \mbox{tr}\,C(R) +\frac{\alpha_s^2}{\pi^2}\Big[ - \frac{1}{2}\, \mbox{tr} \left(C(R)^2\right)+ \frac{3}{2}\, C_2\, \mbox{tr}\,C(R)\nonumber\\&&\times  \Big(\ln a_\varphi + 1 + d_2 - b_{11}\Big) - N_{f} T(R)\,\mbox{tr}\,C(R)\, \Big( \ln a + 1 + d_2 - b_{12} \Big)\Big] + O(\alpha_{s}^3)\biggr),
\noindent
\end{eqnarray}
\noindent
where $b_{11}$ and $b_{12}$ are the finite constants from the one-loop expression for $\alpha_s$ \eqref{One-LoopRunning} and $d_2$ appears in the expression
\begin{equation}\label{EquationForAlpha}
\hspace*{-5mm} \alpha_{0}^{-1} - \alpha^{-1} = -\frac{1}{\pi}\sum\limits_{\alpha=1}^{N_{f}}q_{\alpha}^2 \biggl(\mbox{dim}(R)\Big(\ln\frac{\Lambda}{\mu}+d_{1}\Big) + \frac{\alpha_{s}}{\pi}\, \mbox{tr}\,C(R) \Big(\ln\frac{\Lambda}{\mu}+d_{2}\Big) +O(\alpha_s^2)\biggr).
\end{equation}
\noindent
Similarly, we obtain the anomalous dimension defined in terms of the renormalized coupling constant,
\begin{eqnarray}\label{TwoLoopGammaRenorm}
&& \widetilde{\gamma}(\alpha_{s})_i{}^j = - \frac{\alpha_{s}}{\pi}\, C(R)_i{}^j +\frac{\alpha_s^2}{\pi^2}\Big[ - \frac{3}{2}\, C_2\, C(R)_i{}^j \Big(\ln a_\varphi + 1 + g_{1} - b_{11}\Big)\quad \nonumber\\
&& + N_{f} T(R)\, C(R)_i{}^j  \Big( \ln a + 1 + g_{1} - b_{12}\Big) + \frac{1}{2}\left(C(R)^2\right)_i{}^j\Big] + O(\alpha_{s}^3),
\end{eqnarray}
where $g_1$ is a constant that appears in
\begin{equation}\label{Zed}
\ln Z_i{}^j = \frac{\alpha_{s}}{\pi}\, C(R)_i{}^j \Big(\ln\frac{\Lambda}{\mu} + g_{1}\Big) + O(\alpha_s^2).
\end{equation}
\noindent
From Eqs. \eqref{One-LoopRunning}, \eqref{EquationForAlpha}, and \eqref{Zed} we see that the conditions \eqref{Condition} give $b_{11}=b_{12}=-x_0$, $g_1=-x_0$,  $d_1=d_2=-x_0$. Thus, if the conditions \eqref{Condition} are satisfied, the finite constants cancel each other in Eqs. \eqref{ThreeLoopDRenormalized} and \eqref{TwoLoopGammaRenorm}. Therefore, $\widetilde{D}$ and $\widetilde{\gamma}_i{}^j$ are given by the same functions as $D$ and $\gamma_i{}^j$ and the exact relation \eqref{ShifmanFormula} is valid for them as well.

For $x_0=0$ the conditions \eqref{Condition} require that all the finite constants be equal to zero, and it is only the logarithms that remain (this obviously occurs in all orders). For this reason we call the corresponding prescription $\mbox{HD}+\mbox{MSL}$.

\section*{Acknowledgements}

The authors are grateful to E.A. Ivanov and M.A. Shifman for the  helpful
discussions.
The work of A.L.K. and A.E.K.  was supported by the Foundation for the advancement of theoretical physics and mathematics 'BASIS', grant  No. 17-11-120.


\begin{thebibliography}{10}
\bibitem{Shifman:2014cya}
\textit{Shifman~M., Stepanyantz~K.} Exact Adler Function in Supersymmetric QCD~//Phys. Rev. Lett. 2015. V.~114. P.~051601.

\bibitem{Shifman:2015doa}
\textit{Shifman~M.~A., Stepanyantz~K.~V.} Derivation of the exact expression for the $D$ function in $\mathcal{N}=1$ SQCD~// Phys. Rev. D. 2015. V.~91. P.~105008.

\bibitem{Novikov:1983uc}
\textit{Novikov~V.~A., Shifman~M.~A., Vainshtein~A.~I., Zakharov~V.~I.} Exact Gell-Mann-Low Function of Supersymmetric Yang-Mills Theories from Instanton Calculus~//
Nucl. Phys. B. 1983. V.~229. P.~381.


\bibitem{Kataev:1983at}
\textit{Kataev~A.~L., Pivovarov~A.~A.} {Perturbative corrections to $\sigma_{tot} (e^+ e^- \to hadrons)$ in supersymmetric QCD}~// JETP Lett. 1983. V.~38. P.~369. [Pisma Zh. Eksp. Teor. Fiz. 1983. V.~38. P.~309].

\bibitem{Altarelli:1983pr}
\textit{Altarelli~G., Mele~B., Petronzio~R.} {Broken Supersymmetric QCD and $e^+ e^-$ Hadronic Cross-Sections}~//
Phys. Lett. 1983.  V.~129B. P.~456.

\bibitem{Kataev:2017qvk}
\textit{Kataev~A.~L., Kazantsev~A.~E., Stepanyantz~K.~V.} The Adler $D$-function for $\mathcal{N}=1$ SQCD regularized by higher covariant derivatives in the three-loop approximation~//
Nucl. Phys. B. 2018. V. 926. P.~295.

\bibitem{Kataev:2013eta}
\textit{Kataev~A.~L., Stepanyantz~K.~V.} NSVZ scheme with the higher derivative regularization for $\mathcal{N} =$ 1 SQED~// Nucl. Phys. B. 2013. V.~875. P.~459.




\end{thebibliography}
\end{document}